\documentclass[aps,prl,twocolumn,showpacs,superscriptaddress,floatfix]{revtex4}
\usepackage{here}
\usepackage{graphicx}

\begin{document}


\title{Giant spin canting in the $S$ = 1/2 antiferromagnetic chain [CuPM(NO$_3$)$_2$(H$_2$O)$_2$]$_n$ observed by $^{13}$C-NMR}

\author{A.U.B. Wolter}
\affiliation{Institut f\"{u}r Metallphysik und Nukleare Festk\"{o}rperphysik, TU Braunschweig, 38106 Braunschweig, Germany}
\author{P. Wzietek}
\affiliation{Laboratoire de Physique des Solides, Universit\'{e} Paris-Sud, 91405 Orsay}
\author{S. S\"{u}llow}
\affiliation{Institut f\"{u}r Metallphysik und Nukleare Festk\"{o}rperphysik, TU Braunschweig, 38106 Braunschweig, Germany}
\author{F.J. Litterst}
\affiliation{Institut f\"{u}r Metallphysik und Nukleare Festk\"{o}rperphysik, TU Braunschweig, 38106 Braunschweig, Germany}
\author{A. Honecker}
\affiliation{Institut f\"{u}r Theoretische Physik, TU Braunschweig, 38106 Braunschweig, Germany}
\author{W. Brenig}
\affiliation{Institut f\"{u}r Theoretische Physik, TU Braunschweig, 38106 Braunschweig, Germany}
\author{R. Feyerherm}
\affiliation{Hahn--Meitner--Institut GmbH, 14109 Berlin, Germany}
\author{H.-H. Klauss}
\affiliation{Institut f\"{u}r Metallphysik und Nukleare Festk\"{o}rperphysik, TU Braunschweig, 38106 Braunschweig, Germany}

\date{\today}

\begin{abstract}
We present a combined experimental and theoretical study on copper pyrimidine dinitrate [CuPM(NO$_3$)$_2$(H$_2$O)$_2$]$_n$, a one-dimensional
$S$ = 1/2 antiferromagnet with alternating local symmetry. From the local susceptibility measured by NMR at the three inequivalent carbon sites
in the pyrimidine molecule we deduce a giant spin canting, {\it i.e.}, an additional staggered magnetization perpendicular to the applied
external field at low temperatures. The magnitude of the transverse magnetization, the spin canting of (52$\pm$4)$^\circ$ at 10 K and 9.3 T and
its temperature dependence are in excellent agreement with exact diagonalization calculations.
\end{abstract}


\maketitle

One-dimensional quantum magnets show a rich variety of magnetic ground states, such as spin liquids with quantum critical behavior or gaps in
the spin excitation spectrum ~\cite{haldane83,dender97,stone03}. The ideal $S$ = 1/2 antiferromagnetic Heisenberg chain ($S$ = 1/2 AFHC) with
uniform nearest-neighbor exchange coupling is of particular interest, since it is exactly solvable using the Bethe ansatz equations
~\cite{bethe31,takahashi99,klumper00}. Its ground state is a spin singlet with gapless excitations in contrast to $S$ = 1 Haldane-gap systems.

The ground state properties of the $S$ = 1/2 AFHC are highly sensitive to even small modifications, which often result in real spin chain
systems from a low lattice symmetry. The case of an alternating local environment of the magnetic ion can be treated theoretically including the
Dzyaloshinskii-Moriya (DM) interaction and/or a staggered $g$ tensor, both as consequence of the residual spin-orbit coupling
~\cite{AO,essler,asano00}. Whereas the Heisenberg exchange $J {\bf S}_i {\bf S}_{i+1}$ prefers collinear spin arrangements, the DM interaction
${\bf D} ({\bf S}_i \times {\bf S}_{i+1})$ prefers canted ones. This can be described by the Hamiltonian ~\cite{AO}

\begin{equation}
\hat{H} = J\sum_{i} [{\bf S}_i{\bf S}_{i+1}-h_uS_i^z-(-1)^ih_sS_i^x],
\label{Hop}
\end{equation}

\noindent which includes the effective uniform field $h_u = g\mu_BH/J$ and the induced staggered field $h_s$ $\propto$ $H$ perpendicular to the
applied magnetic field $H$. We refer to this as the {\it staggered} $S$ = 1/2 AFHC model. An essential result is the opening of a spin
excitation gap in external fields, with the excitation spectrum consisting of solitons, antisolitons and their bound states called breathers
~\cite{essler,asano00}.

This model has been used to describe several 1-D spin chain systems, that is, copper benzoate ~\cite{dender97,asano00}, copper pyrimidine
dinitrate [CuPM(NO$_3$)$_2$(H$_2$O)$_2$]$_n$ (CuPM) ~\cite{feyerherm00,wolterrc03,kolezhuk04} and CuCl$_2$ $\cdot$ 2(dimethylsulfoxide)
~\cite{broholm04}. However, generic features of this model, {\it i.e.}, the magnitude and direction of a transverse staggered magnetization
$m_{s\perp}$ and its temperature dependence have not been verified experimentally by now.

In this Letter we report the first direct observation of the staggered magnetization $m_{s\perp}$ in a $S$ = 1/2 linear chain, {\it i.e.} CuPM,
via a detailed $^{13}$C-NMR study. We compare our data with exact diagonalization calculations based upon the staggered $S$ = 1/2 AFHC model. At
10 K and 9.3 T the transverse magnetization gives rise to a giant spin canting of (52$\pm$4)$^\circ$ with respect to the external field. This
observation manifests the strong influence of spin orbit coupling in this system.

We measured the local susceptibility via the NMR frequency shift $\delta$ at three inequivalent carbon sites in the pyrimidine molecule as a
function of temperature and magnetic field orientation. The transverse staggered magnetization $m_{s\perp}$ is identified both {\it (i)} as a
low temperature deviation from the linear correlation between local and macroscopic susceptibility and {\it (ii)} from the orientation
dependence of $\delta$ at a fixed temperature. The observed magnitude ($m_{s\perp}$ $\approx$ 0.13 $\mu_B$ at 10 K in 9.3 T) and its temperature
(T) dependence are in excellent agreement with the staggered $S$ = 1/2 AFHC model.

Single crystals of CuPM have been grown as described previously ~\cite{ishida97}. The Cu ions form uniformly spaced chains parallel to the short
$\it ac$ diagonal of the monoclinic crystal structure. The intrachain magnetic exchange pathway is provided by the pyrimidine ring (Figs. 1 and
2). From a single crystal study of CuPM an exchange parameter $J/k_B$ = 36.3(5) K is derived ~\cite{feyerherm00,wolterrc03}. An additional
Curie-like contribution to the magnetic susceptibility at low temperatures is observed. It varies strongly with magnitude and direction of the
external field and is identified as the longitudinal component $m_{s||}$ of the total staggered magnetization $m_s$. Following Refs.
~\cite{feyerherm00,wolterrc03} we denote with $m_s$ the magnetization induced by the staggered field which has both a staggered ($m_{s\perp}$)
and a uniform ($m_{s||}$) component. In CuPM $m_{s||}$ is only $\leq$ 0.11 of $m_s$ and adds to the uniform magnetization $m_u$ induced by the
external field $H$. The different contributions to the magnetization induced by either the external or staggered field are illustrated in Fig.
1.

\begin{figure}
\begin{center}
\includegraphics[width=0.8\columnwidth]{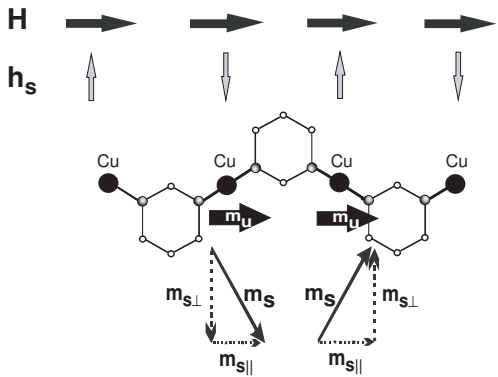}
\end{center}
\caption[1]{A chain segment of [CuPM(NO$_3$)$_2$(H$_2$O)$_2$]$_n$ viewed along the $b$-direction. For clarity, only the Cu ions and the
pyrimidine molecules are shown. The directions of the different components to the magnetization, {\it i.e.}, $m_u$, $m_s$ =
$m_{s||}$+$m_{s\perp}$, are illustrated on the middle chain segment. Note that the different staggered components are not to scale.}
\label{fig:fig1}
\end{figure}

We have performed NMR experiments between 5-200 K in a magnetic field of 9.3 T. The single crystal was oriented with the $ac$-plane parallel to
the external field. The NMR spectra of $^{13}$C have been recorded using a progressive saturation sequence with constant delay and Hahn
spin-echo detection. The NMR shift $\delta$ is defined as the normalized difference between the observed resonance frequency $\omega_{res}$ and
the calculated value for the bare nucleus, $\delta$ = $\frac{\omega_{res}-\gamma \mu_0 H_0}{\gamma \mu_0 H_0}$. $\gamma$ is the gyromagnetic
ratio of the nucleus, $H_0$ was determined from the $^1$H-NMR resonance frequency of water at room temperature. In CuPM, the three inequivalent
carbon sites C1, C2 and C3  with additional hyperfine coupling to the nearest proton ($I$ = 1/2) result in three pairs of resonance lines (Fig.
2) ~\cite{wolterppar03}.

\begin{figure}
\begin{center}
\includegraphics[width=0.8\columnwidth]{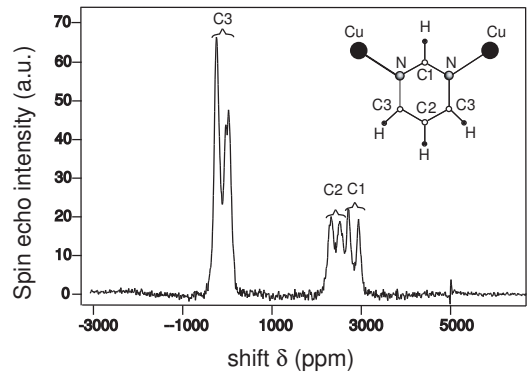}
\end{center}
\caption[1]{A typical $^{13}$C-NMR spectrum of CuPM at 15 K in an external field of 9.3 T applied along the chain direction. The site assignment
of the observed signals has been derived from the angular dependent NMR shift $\delta$ at 200 K ~\cite{wolterppar03}.} \label{fig:fig1}
\end{figure}

\begin{figure}
\begin{center}
\includegraphics[width=1.0\columnwidth]{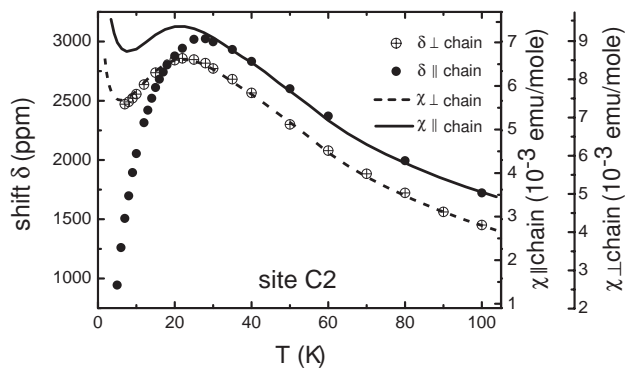}
\end{center}
\caption[1]{The temperature dependence of the NMR shift $\delta$ of CuPM for $H$ $||$ and $\perp$ chain for carbon site C2. The solid and dashed
lines represent the corresponding experimental bulk susceptibility $\chi$. The proportionality constants between the scales for $\delta$ and
$\chi$ give the hyperfine coupling constants $A_{||}$ and $A_{\perp}$ for $H$ $||$ and $\perp$ chain, respectively.} \label{fig:fig1}
\end{figure}

Fig. 3 shows the T dependent shift $\delta$ of CuPM for $H$ $||$ and $\perp$ to the copper chains for carbon site C2 ~\cite{wolterppar03}. Each
set of hyperfine doublets is represented by its average shift. $\delta$ can be described by the sum of the chemical shift $\sigma_0$ and the
Knight shift $K$. While $\sigma_0$ represents the T independent orbital shift due to closed electronic shells, $K$ = $A$$\cdot$$\chi$(T)
describes the hyperfine coupling to the paramagnetic electronic moments mainly residing on the Cu sites. Here, $A$ is the hyperfine coupling
constant, which can either have positive or negative sign, and $\chi$(T) is the magnetic susceptibility. The solid and dashed lines in Fig. 3
represent the experimental bulk susceptibility $\chi$ of CuPM for $H$ $||$ and $\perp$ chains. Clearly, $\chi$(T) is similar to $\delta$(T) for
$T$ $>$ 30 K. However, for $H$ $||$ chain distinct deviations are present for lower temperatures.

A further comparison of the NMR shift $\delta$(T) with $\chi$(T) for all three carbon sites is shown in Fig. 4. Here, the solid lines represent
linear fits of the form $\delta$(T) = $\sigma_0$+$A$$\cdot$$\chi$(T) for $T$ $\ge$ 30 K. Whereas for $H$ $\perp$ chain this linear relation is
obeyed in the full temperature range (5-120 K), a large deviation is observed below 30 K for $H$ $||$ chain. In this geometry the transverse
component of the staggered magnetization $m_{s\perp}$ results in an additional Knight shift $K_{s}$ ~\cite{note3}.

$K_{s}$ is extracted from the data via $K_s$(T) = $\delta$(T)-$\sigma_0$-$A$$\cdot$$\chi$(T) and is shown in Fig. 5 (a) for sites C1, C2 and C3,
respectively. The solid lines represent fits to $K_s$(T) = $A_{dip,\uparrow\downarrow}$$\cdot$$C_s$/T+$K_{s,corr}$. The hyperfine coupling
constant for a staggered magnetization along the $b$-axis, $A_{dip,\uparrow\downarrow}$, is calculated in localized dipole approximation within
a sphere of 120 \AA~centered at the respective carbon site. A small offset $K_{s,corr}$ $\approx$ -250 ppm had to be included since in this
analysis the experimental value $K_s$(30K) is fixed to zero. From the fitted parameters $C_s$ for the three carbon sites we deduce independent
values for $m_{s\perp}$, namely $m_{s\perp}$(C1)=(0.07$\pm$0.01)$\mu_B$, $m_{s\perp}$(C2)=(0.16$\pm$0.01)$\mu_B$ and
$m_{s\perp}$(C3)=(0.07$\pm$0.01)$\mu_B$ at 10 K and $\mu_0H$ = 9.3 T.

In Fig. 5 (b) we compare the average of the experimental results obtained from the three carbon sites, $\overline{m_{s\perp}}$ and ($\overline
{m_u + m_{s||}}$), with results for the uniform $z$-component $m_u$ and the staggered $x$-component $m_s$ of the magnetization obtained by full
diagonalization of the $S=1/2$ AFHC chain Hamiltonian (1) with $N=16$ sites and periodic boundary conditions. Here, we have used $J/k_B$ = 36.5
K, a ratio of the staggered and uniform field $h_s$/$h_u$ = 0.083 and $g$ = 2.117 for $H$ $||$ chain ~\cite{feyerherm00,wolterrc03}. Comparison
with results for $N < 16$ (not shown) and for $N=20$ at $T=0$ ~\cite{wolterrc03} indicates that the data for $N=16$ yield a good approximation
to the thermodynamic limit at all temperatures. We find excellent agreement between experiment and theory in the whole temperature range. At 10
K and 9.3 T $||$ chain the ratio of the staggered magnetization, $\overline {m_{s\perp}}$ = 0.13 $\mu_B$, to the total uniform one, $\overline
{(m_u + m_{s||})}$ = 0.10 $\mu_B$, corresponds to a giant spin canting of (52$\pm$4)$^\circ$ with respect to the external field. With decreasing
temperature the spin canting increases even further, extrapolating to $\sim$ 75$^\circ$.

\begin{figure}
\begin{center}
\includegraphics[width=0.8\columnwidth]{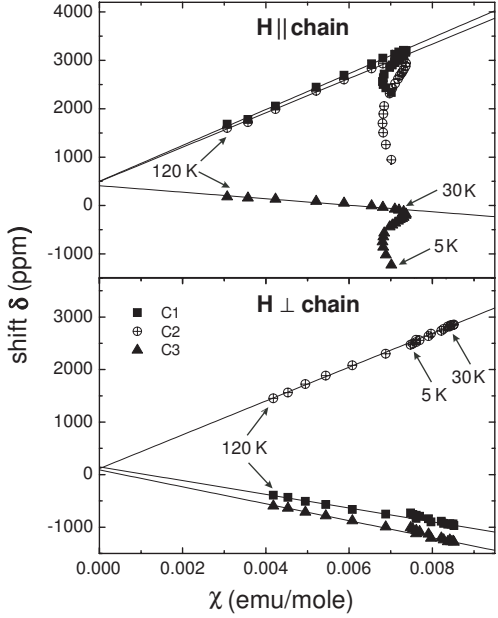}
\end{center}
\caption[1]{NMR shifts $\delta$ vs experimental magnetic susceptibility $\chi$ for CuPM with the external field $H$ $||$ and $\perp$ to the
chain in the {\it ac}-plane. The solid lines are fits of the form $\delta$(T)=$\sigma_0$+$A$$\cdot$$\chi$(T) for $T$ $\ge$ 30K.}
\end{figure}

\begin{figure}
\begin{center}
\includegraphics[width=0.9\columnwidth]{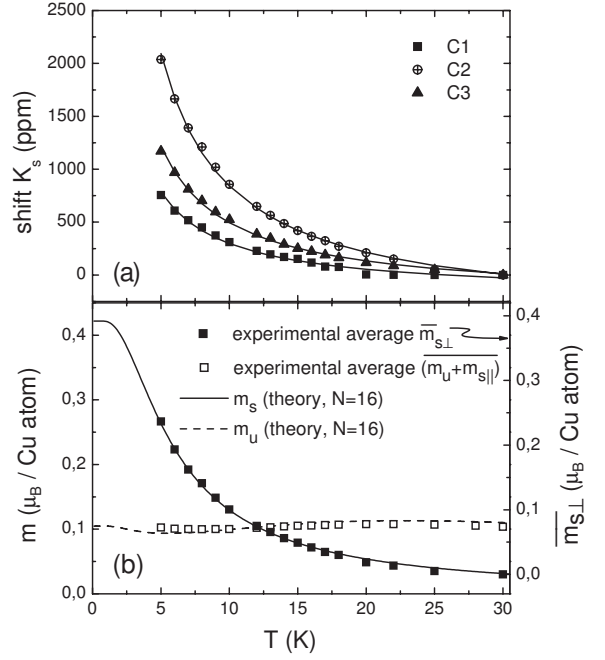}
\end{center}
\caption[1]{(a) The T dependent transverse staggered contribution to the Knight shift, $K_{s}$, in CuPM with $H$ $||$ to the Cu chains. The
solid lines are a parametrization to $K_s$ = $A_{dip,\uparrow\downarrow}$$\cdot$$C_s$/T+$K_{s,corr}$. (b) $\overline{m_{s\perp}}$ and
($\overline {m_u + m_{s||}}$), the experimental average of the three carbon sites of the staggered and total uniform components, respectively,
in comparison with exact diagonalization calculations of a linear chain with $N$ = 16 spins for both, the staggered magnetization $m_s$ and
$m_u$. Note that the small offset (-0.02 $\mu_B$/ Cu atom) of the magnetization scales between the experimental $\overline{m_{s\perp}}$ and
calculated data results from the analysis which sets $K_{s}$(30 K) = 0.} \label{fig:fig4}
\end{figure}

In an independent approach to extract $m_{s\perp}$ we measured the angular dependence of $\delta$ in the {\it ac}-plane at 200 K
~\cite{wolterppar03}, 30 K and 10 K (Fig. 6). We take into account anisotropic dipole and isotropic hyperfine coupling to the longitudinal
susceptibility ${\bf\underline \chi}_u$+${\bf\underline \chi}_{s||}$, anisotropic dipole coupling to the transverse susceptibility
${\bf\underline \chi}_{s\perp}$, and the orbital shift ${\bf\underline \sigma}$:

\begin{eqnarray}
\delta & = & \frac{1}{H^2} {\bf H} \cdot({\bf \underline {A}}_{dip,\uparrow\uparrow} \cdot ({\bf
\underline{\chi}}_{u}+{\bf\underline{\chi}}_{s||}) + A_{iso}({\bf\underline {\chi}}_{u}+{\bf\underline {\chi}}_{s||}) + \nonumber\\ & & +
{\bf\underline {A}}_{dip,\uparrow\downarrow} \cdot {\bf\underline{\chi}}_{s\perp} + {\bf\underline{\sigma}})\cdot {\bf H}. \label{mphys}
\end{eqnarray}

The dipole hyperfine tensors for uniform and staggered susceptibility, ${\bf\underline A}_{dip,\uparrow\uparrow}$ and ${\bf\underline
A}_{dip,\uparrow\downarrow}$, are obtained by dipole field calculations as described above. In order to adequately describe the data one needs
to take into account a finite moment transfer of 10$\%$ to the nitrogen atoms of the pyrimidine molecules in the calculation of ${\bf\underline
A}_{dip,\uparrow\uparrow}$. The moment transfer is consistent with preliminary electronic structure calculations ~\cite{dollpc} and it is close
to the value observed in a molecular magnet system with MnCu antiferromagnetic chains ~\cite{gillon99}. We also considered a finite moment
transfer to the nitrogen atoms in the calculations of ${\bf \underline{A}}_{dip,\uparrow\downarrow}$. Here, a 100$\%$ transverse moment on the
Cu site yields the best description of our data and is used throughout this work. The orbital shift tensor ${\bf\underline{\sigma}}$ was
calculated from $\sigma_0$ measured $||$ and $\perp$ to the chain ~\cite{note4}. For 30 K the contributions from the staggered susceptibilities,
${\bf\underline{\chi}}_{s||}$ and ${\bf\underline{\chi}}_{s\perp}$, are nearly zero. Thus, to describe the experimental data at 30 K, $A_{iso}$
is the only fit parameter and is determined to: $A_{iso,C1}$ = (0.05 $\pm$ 0.01)mole/emu, $A_{iso,C2}$ = (0.38 $\pm$ 0.01)mole/emu and
$A_{iso,C3}$ = (-0.07 $\pm$ 0.01)mole/emu.

Using the isotropic constants $A_{iso}$ determined at T = 30 K and the experimental values for ${\bf\underline{\chi}}_{s||}$ from Ref.
~\cite{feyerherm00}, the remaining parameter to fit our NMR shift data at 10 K is the transverse staggered susceptibility
${\bf\underline{\chi}}_{s\perp}$. The solid lines in Fig. 6(b) represent the fits to Eq. (2). We obtain $m_{s\perp}$(C1) =
(0.09$\pm$0.01)$\mu_B$, $m_{s\perp}$(C2) = (0.20$\pm$0.02)$\mu_B$ and $m_{s\perp}$(C3) = (0.06$\pm$0.01)$\mu_B$ for $H$ $||$ chain. These values
are fully consistent with the results of the analysis of the temperature dependence of $K_{s}$, as presented above. If the transverse staggered
component ${\bf H}\cdot{\bf\underline{A}}_{dip,\uparrow\downarrow}\cdot{\bf\underline{\chi}}_{s\perp}\cdot{\bf H}$ is omitted in our description
(dashed curves in Fig. 6(b)), our data cannot be reproduced.

We believe that the variance of the results of $m_{s\perp}$ for the three inequivalent carbon sites stems from the localized dipole
approximation which we used to calculate the dipole hyperfine coupling tensors ${\bf\underline A}_{dip}$. This indicates that in order to
improve the description of our data the effect of delocalization of spin-density ought to be considered by means of extended electronic
structure calculations.

\begin{figure}
\begin{center}
\includegraphics[width=0.8\columnwidth]{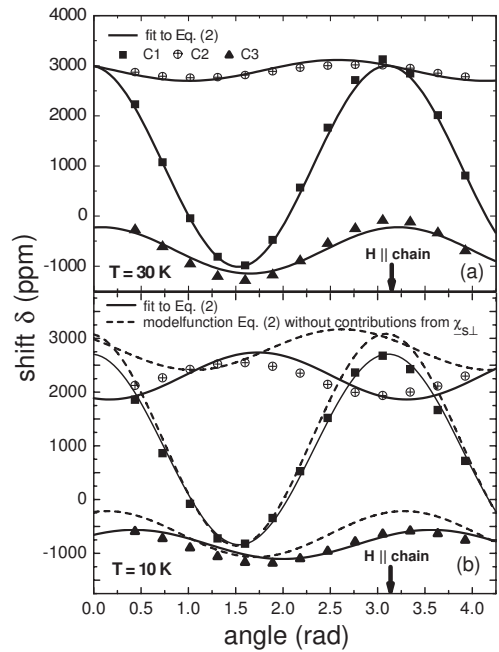}
\end{center}
\caption[1]{The angular-dependent NMR shift $\delta$ of CuPM at $T$ = 30 K and 10 K, respectively, with the external field aligned in the
$ac$-plane. For further details see text.} \label{fig:fig5}
\end{figure}

In conclusion, we have performed $^{13}$C-NMR experiments on [CuPM(NO$_3$)$_2$$\cdot$(H$_2$O)$_2$]$_n$ as function of temperature and magnetic
field orientation. We extracted the transverse staggered magnetization as a low temperature deviation from the linear correlation between local
and macroscopic susceptibility, and from the orientation dependence of the NMR frequency shift at 10 K. The observed temperature dependence is
in excellent agreement with theoretical results for the staggered $S$ = 1/2 AFHC model. The observed giant spin canting highlights the strong
influence of only weak residual spin orbit interactions in such systems. Our data also provide detailed information on the hyperfine coupling in
[CuPM(NO$_3$)$_2$(H$_2$O)$_2$]$_n$ as well as on the absolute value of the staggered magnetization.

This work has partially been supported by the DFG under Contract No. KL1086/4-2 and the European Community Marie-Curie fellowship programme
STROCOLODI. A.U.B. Wolter would like to thank the Laboratoire de Physique des Solides for hospitality and D. J\'{e}rome for fruitful
discussions. The numerical results presented in Fig. 4(b) were obtained on the compute-server {\tt cfgauss} of the TU Braunschweig.


\begin{references}
\bibitem{haldane83} F.D.M. Haldane, Phys. Rev. Lett. {\bf 50}, 1153 (1983).
\bibitem{dender97} D.C. Dender {\it et al.}, Phys. Rev. Lett. {\bf 79}, 1750 (1997).
\bibitem{stone03} M.B. Stone {\it et al.}, Phys. Rev. Lett. {\bf 91}, 037205 (2003).
\bibitem{bethe31} H.A. Bethe, Z. Phys. {\bf 71}, 205 (1931).
\bibitem{takahashi99} M. Takahashi, {\it Thermodynamics of One-Dimensional Solvable Models} (Cambridge University Press, Cambridge, 1999).
\bibitem{klumper00} A. Kl\"{u}mper and D.C. Johnston, Phys. Rev. Lett. {\bf
    84}, 4701 (2000).
\bibitem{AO} M. Oshikawa and I. Affleck, Phys.\ Rev.\ Lett.\ {\bf 79}, 2883 (1997); I. Affleck and M. Oshikawa, Phys. Rev. B {\bf 60}, 1038
  (1999).
\bibitem{essler} F.H.L. Essler and A.M. Tsvelik, Phys. Rev. B {\bf 57}, 10592
  (1998); F.H.L. Essler, Phys. Rev. B {\bf 59}, 14376 (1999).
\bibitem{asano00} T. Asano {\it et al.}, Phys. Rev. Lett. {\bf 84}, 5880 (2000).
\bibitem{feyerherm00} R. Feyerherm {\it et al.}, J. Phys.:
  Condens. Matter {\bf 12}, 8495 (2000).
\bibitem{wolterrc03} A.U.B. Wolter {\it et al.}, Phys. Rev. B {\bf
68}, 220406(R) (2003).
\bibitem{kolezhuk04} S.A. Zvyagin {\it et al.}, Phys. Rev. Lett. {\bf 93}, 027201 (2004) .
\bibitem{broholm04} M. Kenzelmann {\it et al.}, Phys. Rev. Lett. {\bf 93}, 017204 (2004).
\bibitem{ishida97} T. Ishida {\it et al.}, Synth. Metals {\bf 85}, 1655 (1997).
\bibitem{wolterppar03} A.U.B. Wolter {\it et al.}, Polyhedron {\bf 22}, 2273 (2003).
\bibitem{note3} Although for $H$ $\perp$ chain a small transverse staggered
magnetization $m_{s\perp}$ is also present ~\cite{feyerherm00}, the dipole coupling constant for a transverse spin polarisation is nearly zero
in this geometry and thus one cannot detect an additional Knight shift $K_s$.
\bibitem{dollpc} K. Doll, private communication.
\bibitem{gillon99} B. Gillon, Mol. Cryst. and Liq. Cryst. {\bf 335},
53 (1999).
\bibitem{note4}Assuming the principal axes of the
orbital shift tensor {\bf \underline{$\sigma$}} to be the C-H bond
axis, the perpendicular axis lying in the plane of the pyrimidine
molecule and the axis perpendicular to the pyrimidine ring, we
obtain the two diagonal elements of {\bf \underline{$\sigma$}}
which are necessary for a rotation of a magnetic field $H$ in the
$ac$-plane.
\end{references}
\end{document}